\documentstyle[12pt]{article}
\begin{document}
\def\t{\times}\def\p{\phi}\def\P{\Phi}\def\a{\alpha}\def\e{\varepsilon}
\def\be{\begin{equation}}\def\ee{\end{equation}}\def\l{\label}
\def\0{\setcounter{equation}{0}}\def\b{\beta}\def\S{\Sigma}\def\C{\cite}
\def\r{\ref}\def\ba{\begin{eqnarray}}\def\ea{\end{eqnarray}}\def\n{\nonumber}
\def\R{\rho}\def\X{\Xi}\def\x{\xi}\def\La{\Lambda}\def\la{\lambda}
\def\d{\delta}\def\s{\sigma}\def\f{\frac}\def\D{\Delta}\def\pa{\partial}
\def\Th{\Theta}\def\o{\omega}\def\O{\Omega}\def\th{\theta}\def\ga{\gamma}
\def\Ga{\Gamma}\def\h{\hat}\def\rar{\rightarrow}\def\vp{\varphi}
\def\inf{\infty}\def\le{\left}\def\ri{\right}\def\foot{\footnote}
\def\un{\underline}\def\ve{\varepsilon}\def\po{\propto}\def\Re{\rm
Re}\def\Im{\rm Im}

\begin{center}
{\huge\bf Soft partons production in deep inelastic kinematics}

\vskip 0.3cm
{\large\it J.Manjavidze,~A.Sissakian}\\
{large JINR, Dubna}
\end{center}
\vskip 1cm
\begin{abstract}
It is argued considering multiple production in the deep inelastic
scattering kinematics that the very high multiplicity events are
extremely sensitive to the low-x partons density.

\end{abstract}

\section{Introduction}

The role of soft color partons in the high energy hadron interactions
is the mostly intriguing modern problem of particles physics. So, the
collective phenomena and symmetry breaking in the non-Abelian gauge
theories, confinement of colored charges and the infrared divergences
of the pQCD are the phenomena just of the soft color particles
domain.

It seems natural that the very high multiplicity (VHM) hadron
interaction, where the energy of created particles is small, should
be sensitive to the soft color particles densities. Indeed, the aim
of this paper is to show that even in the hard by definition deep
inelastic scattering (DIS) the soft color particles role becomes
important in the VHM region.

The standard (mostly popular) hadron theory considers perturbative QCD
(pQCD) at small distances (in the scale of $\La\simeq 0.2~Gev$) as the
exact theory. This statement is confirmed by a number of experiments,
e.g. the DIS data, the QCD jets observation. But one should have in
mind that the pQCD has finite range of validity since the
non-perturbative effects should be taken into account at distances
larger then $1/\La$.

It is natural to assume building the complete theory that at large
distances the non-perturbative effects exceed the perturbative
ones\foot{The corresponding formalism was described e.g. in
\C{man}.}. In result the pQCD contributions becomes $screened$ by
the non-perturbative interactions.

But exist another, more natural, possibility.  It should be
noted here that the pQCD running coupling constant
$\a_s(q^2)=1/b\ln(q^2/\La^2)$ becomes infinite at $q^2=\La^2$ and we
do not know what happens with pQCD if $q^2<\La^2$. There is a
suspicion \C{callan} that at $q^2\sim \La^2$ the properties of
theory changed so drastically (being defined on new vacuum) that even
the notions of pQCD is $disappeared$. This means that pQCD should be
truncated from below on the `fundamental'  scale $\La$. It seems
evident that this infrared cut-off should influence on the soft
hadrons emission in the VHM domain.

So, it is important try to rise predictability of pQCD in the
`forbidden' area of large distances.  For this purpose one should split
experimentally  the perturbative and non-perturbative effects at the
large distances to check out quantitatively the role of soft color
partons. One must realize for this purpose highly unusual condition
that the non-perturbative effects must be negligible even if the
distance among color charges is high.

The non-perturbative effects lead to  the strong polarization of
QCD vacuum and, in result, to the color charge confinement. This
vacuum is unstable \C{casher} against creation of real quark-antiquark
$(q\bar q)$ pairs if the distance among charges became
large. This pairs are captured into the colorless hadrons and just
emission of this `vacuum' hadrons is the mostly important
non-perturbative effect. Therefore, generally, the number of hadrons
$n$, produced even in the hard DIS process,  $n\neq n_{p}$, where
$n_{p}$ is the number of $\bar{q}q$ pairs created `perturbatively'.

Notice, if the kinetic energy of colored partons is small, i.e. is
comparable with hadron masses, creation of `vacuum' hadrons should be
negligible. Just this is the VHM  process kinematics: because of the
energy-momentum conversation law, produced (final-state) partons can
not have high relative momentum and, if they was created at small
distances, the production of `vacuum' hadrons will be negligible (or
did not play important role). Therefore, if the `vacuum' channel is
negligible, the pQCD contributions only should be considered \C{siss}.

The aim of this article is to show that $n\to\infty$ unavoidably
leads to `low-$x$' domain.

\section{DIS kinematics in the VHM domain}\0

To describe the hadron production in pQCD terms the parton-hadron
duality is assumed. This means that the multiplicity, momentum
etc. distributions of hadron and colored partons are the same. This
reduce the problem practically on the level of QED.

Let us consider now $n$ particles (gluons) creation the DIS. We would
like to calculate $D_{ab}(x,q^2;n)$, where
\be
\sum_n D_{ab}(x,q^2;n)= D_{ab}(x,q^2).
\l{1}\ee
As usual, let $D_{ab}(x,q^2)$ be the probability to find parton $b$
with virtuality $q^2<0$ in the parton $a$ of $\sim\la$ virtuality,
$\la>>\La$ and $\a_s(\la)<<1$.  We always may chose $q^2$ and $x$ so
that the leading logarithm approximation (LLA) will be acceptable.
One should assume also that $(1/x)>>1$ to have the phase space, into
which the particles are produced, sufficiently large.

Then $D_{ab}(x,q^2)$ is described by ladder diagrams. From qualitative
point of view this means approximation of Markovian process of random
walk over coordinate $\ln(1/x)$ and time is $\ln\ln|q^2|$. LLA means
that the `mobility' $\sim \ln(1/x)/\ln\ln\le|q^2\ri|$ should be large
\be
\ln(1/x)>>\ln\ln \le|q^2/\la^2\ri|.
\l{2}\ee
But, on other hand \C{lla},
\be
\ln(1/x)<<\ln\le|q^2/\la^2\ri|.
\l{3}\ee

The leading, able to compensate smallness of $\a_s(\la)<<1$,
contributions give integration over wide range $\la^2<<k_i^2<<-q^2$,
where $k_i^2>0$ is the `mass' of real, i.e.  time-like, gluon.  If
the time needed to capture the parton into the hadron is $\sim(1/\La)$
then the gluon should decay if $k_i^2>>\la^2$.  This leads to
creation of (mini)jets. The mean multiplicity $\bar{n}_j$ in the QCD
jets is high if the gluon `mass' $|k|$ is high: $\ln\bar{n}_j \simeq
\sqrt{\ln(k^2/\la^2)}$.

Rising multiplicity may (i) rise number of (mini)jets $\nu$ and/or
(ii) rise the mean value mass of (mini)jets $<|k_i|>$. We will see
that the mechanism (ii) would be favorable. This is consequence of
the Markovian character of considered process.

But rising mean value of gluon masses $|k_i|$ decrease the range of
integrability over $k_i$, i.e. violate the condition (\r{2}) for
fixed $x$. One can remain the LLA taking $x\to0$. But this may
contradict to (\r{3}), i.e. in any case the LLA becomes invalid in
the VHM domain and the next to leading order corrections should be
taken into account.

Noting that the LLA gives main contribution, that the rising
multiplicity leads to the infrared domain, where the soft gluons
creation becomes dominant.

\section{Preliminary notes}\0

First of all, neglecting the vacuum effects, we introduce definite
uncertainty to the formalism. It is reasonable to define the level of
strictness of our computations. Let us introduce for this purpose the
generating function $T_{ab}(x,q^2;z)$:
\be
D_{ab}(x,q^2;n)=\f{1}{2\pi i}\oint\f{dz}{z^{n+1}}T_{ab}(x,q^2;z).
\l{4}\ee
Having large $n$ the integral may be calculated by the saddle point
method. The smallest solution $z_c$ of the equation
\be
n=z\f{\pa}{\pa z}\ln T_{ab}(x,q^2;z)
\l{5}\ee
defines the asymptotic over $n$ behavior:
\be
D_{ab}(x,q^2;n)\propto e^{-n\ln z_c(x,q^2;n)}.
\l{6}\ee
Using the statistical interpretation of $z_c$ as the fugacity it is
natural to write:
\be
\ln z_c(x,q^2;n)=\f{C_{ab}(x,q^2;n)}{\bar{n}_{ab}(x,q^2)}.
\l{7}\ee
Notice that the solution of eq.(\r{5}) $z_c(x,q^2;n)$ should be the
increasing function of $n$. At first glance this follows from
positivity of all $D_{ab}(x,q^2;n)$. But actually this assumes that
$T_{ab}(x,q^2;z)$ is a regular function of z at $z=1$. This is a
natural assumption considering just the pQCD predictions.

Therefore,
\be
D_{ab}(x,q^2;n)\propto
e^{-\f{n}{\bar{n}_{ab}(x,q^2)}C_{ab}(x,q^2;n)}.
\l{8}\ee
This form of $D_{ab}(x,q^2;n)$ is useful since usually
$C_{ab}(x,q^2;n)$ is the slowly varying  function of $n$. So, for
Poisson distribution $C_{ab}(x,q^2;n)\sim\ln n$. For KNO scaling
we have $C_{ab}(x,q^2;n)=const.$ over $n$.

We would like to note, that neglecting effects of vacuum polarization
we introduce into the exponent so high uncertainty assuming $n\simeq
n_p$ that it is reasonable perform the calculations with the
exponential accuracy. So, we would calculate
\be
-\bar\mu_{ab}(x,q^2;n)=\ln\f{D_{ab}(x,q^2;n)}{D_{ab}(x,q^2)}=
\f{n}{\bar{n}_{ab}(x,q^2)}C_{ab}(x,q^2;n)(1+O(1/n))
\l{9}\ee

The $n$ dependence of $C_{ab}(x,q^2;n)$ defines the asymptotic
behavior of $\bar\mu_{ab}(x,q^2;n)$ and calculation of its
explicit form would be our aim.

\section{Correlation functions}\0

Considering particles creation in the DIS processes one should
distinguish correlation of particles in the (mini)jets and the
correlations between (mini)jets. We will describe the jet
correlations. One should introduce the $\nu$ jets creation
inclusive cross section $\P_\nu^{(r)_\nu}(k_1,k_2,....,k_\nu;q^2,x)$,
where $k_i$, $i=1,2,...,n$ are the jets 4-momentum. Having
$\P_\nu$ we can find the correlations function
$N_\nu^{(r)}(k_1,k_2,....,k_\nu;q^2,x)$, where $(r)=r_1,...,r_\nu$
and $r_i=(q,g)$ defines the sort of created color particle. It is
useful introduce the generating functional
\be
F^{ab}(q^2,x;w)=\sum_n\int d\O_n(k)\prod_{i=1}^n w^{r_i}(k_i)
\le|a^{ab}_n(k_1,k_2,....,k_n;q^2,x)\ri|^2,
\l{10}\ee
where $a^{ab}_n$ is the amplitude, $d\O_n(k)$ is the phase space
volume and $w^{r_i}(k_i)$ are the arbitrary functions. It is evident,
\be
F^{ab}(q^2,x;w)\le.\ri|_{w=1}=D^{ab}(q^2,x).
\l{11}\ee
The inclusive cross sections
\be
\P_\nu^{(r)}(k_1,k_2,....,k_\nu;q^2,x)=\prod_{1=1}^\nu
\f{\d}{\d w^{r_i}(k_i)}F^{ab}(q^2,x;w)\le.\ri|_{w=1}.
\l{12}\ee
The correlation function
\be
N_\nu^{(r)}(k_1,k_2,....,k_\nu;q^2,x)=\prod_{1=1}^\nu
\f{\d}{\d w^{r_i}(k_i)}\ln F^{ab}(q^2,x;w)\le.\ri|_{w=1}.
\l{13}\ee
We will use this definitions to find the partial structure functions
$D^{ab}(q^2,x;n)$.

It will be useful to introduce the Laplace transform over variable
$\ln(1/x)$:
\be
F^{ab}(q^2,x;w)=\int_{\Re j<0}\f{dj}{2\pi i}(\f{1}{x})^j
f^{ab}(q^2,j;w)
\l{14}\ee

The expansion parameter of our problem $\a_s\ln(-q^2/\la^2)\sim1$. By
this reason one should take into account all possible cuttings of the
ladder diagrams. So, calculating $D^{ab}(q^2,x)$in the LLA all
possible cuttings of sceleton ladder diagrams is defined by the
factor \C{lla}:
\be
\f{1}{\pi}\le\{\Ga^{ab}_{r}G_{r}\Ga_{r}^{ab}\ri\},
\l{15}\ee
i.e. the cutting line may get not only through the exact Green
function $G_{r}(k^2_i)$ but through the exact vertex functions
$\Ga^{ab}_{r}(q_i,q_{i+1},k_i)$ also ($q_i^2,q_{i+1}^2$ are
negative). We have in the LLA that
$$
\la^2<<-q^2_i<<-q_{i+1}^2<<-q^2
$$
and
$$
x\leq x_{i+1}\leq x_i\leq 1.
$$
Following to our approximation, see previous section, we would not
distinguish the way as cut line go through the Born amplitude
$$
a^{ab}_{r})=\le\{(\Ga^{ab}_{r})^2G_{r}\ri\}.
$$
We will simply associate $w^{r}\Im a^{ab}_{r})$ to each rung of the
ladder.

Considering the asymptotics over $n$ the time-like partons virtuality
$k_i\simeq -q^2_i/y_i$ should be maximal. Here $y_i$ is the fraction
of the longitudinal momentum of the jet. Then, slightly limiting the
jets phase space,
\be
\ln k_i^2=\ln \le|q_{i+1}\ri|^2(1+O(\ln(1/x)/|q_{i+1}|^2)).
\l{16}\ee

In result, introducing useful in the LLA variable
$\tau_i=\ln(q^2_i/\La^2)$, where $\a_s(q^2)=1/\b\tau$,
$\b=(11N/3)-(2n_f/3)$, we can find following set of equations:
\be
\tau\f{\pa}{\pa\tau}f_{ab}(q^2,j;w)=\sum_{c,r}\vp^r_{ac}(j)w^r(\tau)
f_{ab}(q^2,x;w),
\l{17}\ee
where
\be
\vp^r_{ac}(j)=\vp_{ac}(j)=\int^1_0\f{dx}{x}x^jP_{ac}(x)
\l{18}\ee
and $P_{ac}(x)$ is the regular kernel of the Bethe-Salpeter equation
\C{lla}. At $w=1$ this equation is the ordinary one for
$D^{ab}(q^2,x)$.

We will search the correlation functions from eq.(\r{17}) in terms of
Laplace transform $n_{ab}^{(r)_\nu}(k_1,k_2,....,k_\nu;q^2,j)$. Let
us write:
\be
f_{ab}(q^2,j;w)=d_{ab}(q^2,j)\exp\le\{
\sum_\nu\f{1}{\nu!}\int\prod_{i=1}^\nu\le(\f{d\tau_i}{\tau_i}
(w^{r_i}(\tau_i)-1)\ri)n_{ab}^{(r)_\nu}(k_1,k_2,....,k_\nu;q^2,j)
\ri\}
\l{19}\ee
Inserting (\r{19}) into (\r{17}) and expanding over $(w-1)$ we find
the sequence of coupled equation.

Omitting the cumbersome calculations, we write in the LLA that
\be
\p_{ab}^{(r)_\nu}(\tau_1,\tau_2,....,\tau_\nu;q^2,j)=
d_{ac_1}(j,\tau_1)\vp^{r_1}_{c_1c_2}(j)d_{c_2c_3}(j,\tau_2)\cdots
\vp^{r_\nu}_{c_\nu c_{\nu+1}}(j)d_{c_{\nu+1}b}(j,\tau_{\nu+1}).
\l{20}\ee
One should take into account the conservation laws:
\be
\tau_1\cdot\tau_2\cdots\tau_{\nu+1}=\tau,~
\tau_1<\tau_2<\dots<\tau_{\nu+1}<\tau.
\l{21}\ee
Computing the Laplace transform of this expression we find
$\P_{ab}^{(r)_\nu}(\tau_1,\tau_2,....,\tau_\nu;q^2,x)$.

The kernel $d_{ab}(j,\tau)$ was introduced in (\r{20}). Let us write
it in the form:
\be
d_{ab}(j,\tau)=\sum_{\s=\pm}\s\f{d_{ab}(j)}{\nu_+-\nu_-}\tau^{\nu_\s(j)},
\l{22}\ee
where
\be
d^\s_{qq}=\nu_s-\vp_{gg},~d^\s_{qg}=\nu_s-\vp_{qq},~
d^\s_{qg}=\vp_{gq},~d^\s_{gq}=\vp_{qg}
\l{23}\ee
and
\be
\nu_\s=\f{1}{2}\le\{\vp_{qq}+\vp_{gg}+\s\le[(\vp_{qq}-\vp_{gg})^2-
4n_f\vp_{qg}\vp_{gq}\ri]^{1/2}\ri\}.
\l{24}\ee
If $x<<1$, then $(j-1)<<1$ are essential. In this case \C{lla},
\be
\vp_{gg}\sim\vp_{gq}>>\vp_{qg}\sim\vp_{qq}=O(1).
\l{25}\ee
This means the gluon jets dominance and
\be
n^g_{gg}=\vp_{gg}+O(1).
\l{26}\ee
One can find following estimation of the two-jet correlation
function:
\be
n_{ab}^{r_1r_2}(\tau_1,\tau_2;\j,\tau)=
O\le({\rm max}\{(\tau_1/\tau)^{\vp_{gg}},(\tau_2/\tau)^{\vp_{gg}},
(\tau_1/\tau_2)^{\vp_{gg}}\}\ri\}.
\l{27}\ee
This correlation function is small since in the LLA
$\tau_1<\tau_2<\tau$. This means that the jet correlations becomes
high iff the mass of correlated jets are comparable. But this
condition shrinks the range of integration over $\tau$ and by this
reason one may neglect the `short-range' correlations among jets.
Therefore, as follows from (\r{19}),
\be
f_{ab}(q^2,j;w)=d_{gg}(\tau,j)\exp\le\{
\vp_{gg}\int^\tau_{\tau_0}\f{d\tau'}{\tau'}
w^{g}(\tau')\ri\}
\l{28}\ee
We will use this expression to find the multiplicity distribution in
the DIS domain.

\section{Generating function in the DIS kinematics}\0

To describe particles production one should replace $w^{r}\Im
a^{ab}_{r})$ on $w^{r}_n\Im a^{ab}_{r})$, where $w^{r}_n$ is the
$probability$ of $n$ particles production,
\be
\sum_nw^{r}_n=1.
\l{29}\ee
Having $\nu$ jets one should take into account the conservation
condition $n_1+n_2+...+n_\nu=n$. By this reason the generating
functions formalism is useful. In result one can find that if we take
(\r{28})
\be
w^g=w^g(\tau,z),~w^g(\tau,z)\le.\ri|_{z=1}=1,
\l{30}\ee
then $f_{ab}(q^2,j;w)$ defined by (\r{28}) is the generating
functional of the multiplicity distribution in the `$j$
representation'. In this expression $w^g(\tau,z)$ is the generating
function of the multiplicity distribution in the jet of mass $|k|=\la
e^{\tau/2}$.

In result, see (\r{14}),
\be
F^{ab}(q^2,x;w)\propto\int_{\Re j<0}\f{dj}{2\pi i}({1}/{x})^j
e^{\vp_{gg}\o(\tau,z)}
\l{31}\ee
where
\be
\o(\tau,z)=\int^\tau_{\tau_0}\f{d\tau'}{\tau'}w^{g}(\tau',z).
\l{32}\ee
Noting the normalization condition (\r{30}),
\be
\o(\tau,z=1)=\ln\tau.
\l{33}\ee

The integral (\r{31}) may be calculated by steepest descent method.
It is not hard to see that
\be
j\simeq j_c=1+\le\{4N\o(\tau,z)/\ln(1/x)\ri\}^{1/2}
\l{34}\ee
is essential. Notice that $j-1<<1$ should be essential we find,
instead of the constraint (\r{2}), that
\be
\o(\tau,z)<<\ln(1/x).
\l{35}\ee
In the frame of this constraint,
\be
F^{ab}(q^2,x;w)\propto \exp\le\{4\sqrt{N\o(\tau,z)\ln(1/x)}\ri\}.
\l{36}\ee

Generally speaking, exist such values of $z$ that $j_c-1\sim1$. This
is possible if $\o(\tau,z)$ is a regular function of $z$ at $z=1$.
Then $z_c$ should be the increasing function of $n$ and consequently
$\o(\tau,z_c)$ would be the increasing function of $n$. Therefore,
one may expect that in the VHM domain $j_c-1\sim1$.

Then $j\simeq 1+\o(\tau,z)/\ln(1/x)$ would be essential in the
integral (\r{31}). This leads to following estimation:
$$
F^{ab}(q^2,x;w)\propto e^{-\o(\tau,z)}.
$$
But this is impossible since $F^{ab}(q^2,x;w)$ should be the
increasing function of $z$. This shows that the estimation (\r{36})
has a finite range of validity.

Solution of this problem with unitarity is evident. One should take
into account correlations among jets considering the expansion
(\r{19}). Indeed, smallness of $n_{ab}^{(r)_\nu}$ may be compensated
by large values of $\prod_i^\nu w^{r_i}(\tau_i,z)$ in the VHM domain.

\section{Conclusion}\0

We can conclude that our LLA is applicable in the VHM domain till
\be
\o(\tau,z)<<\ln(1/x)<<\tau=\ln(-q^2/\la).
\l{37}\ee
The mean multiplicity of gluons created in the DIS kinematics
\be
\bar{n}_g(\tau,x)=\f{\pa}{\pa z}\ln F^{ab}(q^2,x;w)\le.\ri|_{z=1}
=\o_1(\tau)\sqrt{4N\ln(1/x)/\ln\tau}>>\o_1(\tau),
\l{38}\ee
where
\be
\o_1(\tau)=\int^\tau_{\tau_0}\f{d\tau_1}{\tau_1}\bar{n}_j(\tau)
\l{39}\ee
and the mean gluon multiplicity in the jet $\bar{n}_j(\tau)$
has following estimation \C{12}:
\be
\ln\bar{n}_j(\tau)\simeq \sqrt{\tau}
\l{40}\ee
Inserting (\r{40}) into (\r{39}),
$$
\o_1(\tau)=\bar{n}_j(\tau)/\sqrt{\tau}.
$$
Therefore, noting (\r{3}),
\be
\bar{n}_g(\tau,x)\simeq\bar{n}_j(\tau)
\sqrt{4N\ln(1/x)/\tau\ln\tau}<<\bar{n}_j(\tau).
\l{41}\ee
This means that the considered `t-channel' ladder is important in the
narrow domain of multiplicities
\be
n\sim \bar{n}_g<<\bar{n}_j.
\l{42}\ee

So, in the VHM domain $n>>\bar{n}_g$ one should consider\\
(i) The ladder diagrams with small number of rungs;\\
(ii) To take into account the malty-jet correlations assuming that
increasing multiplicity leads to the increasing number of rungs in
the ladder diagram.\\
To choose one of this possibilities one should consider the structure
of $\o(\tau,z)$ much more carefully. This will be done in the
consequent paper.

We can conclude that the VHM domain multiplicities production
unavoidably destroy the ladder LLA. To conserve this leading
approximation one should choose $x\to0$ and, in result, to get to the
multi-ladder diagrams, since in this case $\a_s\ln (-q^2/\la^2)\sim
1$ and $\a_s\ln(1/x)\sim 1$. Such theory was considered in \C{grib}.

\vskip 0.3cm
{\bf Acknowledgments}
\vskip 0.2cm

We are grateful to V.G.Kadyshevski for interest to discussed in the
paper questions. We would like to note with gratitude that the
conversations with E.Kuraev was always interesting and important.

\end{document}